\documentclass{ptapap}

\author{Ennio Poretti}[OAB,TNG,GEOS]
\author{Jean-Francois Le Borgne}[IRAP,CNRS,GEOS]
\author{Alain Klotz}[IRAP,CNRS,GEOS]
\author{Monica Rainer}[OAB,OAFi]
\author{Mercedes Correa}[AAS,GEOS]
\affil[OAB]{INAF -- Osservatorio Astronomico di Brera, \\
Via E. Bianchi 46, 23807 Merate (LC), Italy}\\
\affil[TNG] {INAF -- Fundaci\'{o}n Galileo Galilei, Rambla Jos\'{e} Ana Fernandez P\'{e}rez 7, 38712 -- Bre\~{n}a Baja, Spain}
\affil[GEOS]
{Groupe Europ\'een d'Observations Stellaires (GEOS), Bailleau l'Ev\'eque, France}
\affil[IRAP]
{Universit\'e de Toulouse, UPS-OMP, IRAP, Toulouse, France}
\affil[CNRS]
{CNRS, IRAP, Toulouse, France}
\affil[OAFi]
{INAF -- Osservatorio Astrofisico di
Arcetri, Largo E. Fermi 5, 50125 Firenze, Italy}
\affil[AAS]
{Agrupaci\'o Astron\`omica de Sabadell, Sabadell, Spain}

\title{Cyclic variations in the periods of RR Lyr stars}

\begin{document}

\maketitle

\begin{abstract}

We report here on  two types of cyclic variations that can be observed 
in the periods of
RR~Lyr stars, i.e., the Blazhko and the light-time effects. The former has been
investigated  
 by studying the amplitude variations recorded in RR Lyr itself,
firstly by {\it Kepler} and then by the network of the ``Very Tiny Telescopes" (VTTs).
The latter on the basis of the new spectroscopic observations of the most
promising candidate, KIC~2831097. The start of the search for binary candidates
in the RR Lyr stars observed with the TAROT telescopes is also announced.
\end{abstract}

\section{Introduction}

The RR Lyr variables are subjected to many other periodicities overlapped to the
main pulsation cycle.  The evolutionary changes, the excitation of other radial 
and nonradial modes, the modulation in amplitude and phase known as Blazhko effect 
and the period doubling are  extensively reported in literature. Unfortunately, 
many theoretical aspects are not well understood yet. Recently, many efforts have
been made to detect light-time effect, i.e., periodic variations of the O-C 
({\it Observed} minus {\it Calculated}) values of the times of maximum brightness
($T_{\rm max}$) due to the orbital motion of the RR Lyr component in a binary
system.  

In the present contribution we report on the new observational efforts
made jointly by the Brera team in Italy, the IRAP team in France, and by the 
amateur astronomers composing the {\it  Groupe Europ\'een d'Observations Stellaires}
(GEOS).

\section{The pulsation period of RR Lyr}
\begin{figure}
\begin{minipage}{1.0\textwidth}
\centering
\includegraphics[width=\textwidth]{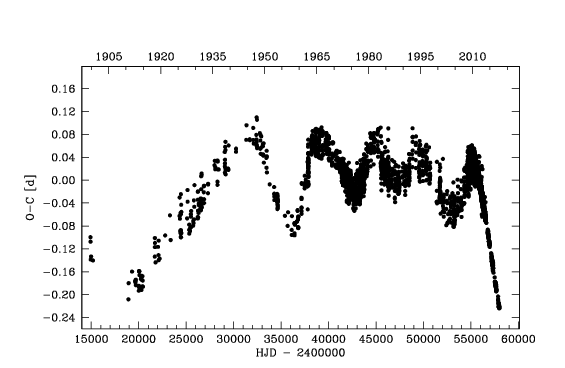}
\caption{The historical behaviour of the O-C values of RR Lyr. The  new $T_{\rm max}$ measured
 by VTTs in 2015, 2016 and 2017 are on the extreme right side.
%shown in red and blue, respectively. 
}
\label{storia} 
\end{minipage}
\end{figure}

The analysis of all the $T_{\rm max}$  epochs listed in the GEOS database   
allowed the careful
 reconstruction of the changes in the pulsation period of RR~Lyr \citep{itself}, 
i.e., the alternation of two states
characterized by the pulsation over a period  longer than 0.56684~d and over another 
shorter than 0.56682~d. 
We emphasize the importance of continuing to
monitor RR~Lyr: it is not only the eponym of the
class of variable stars, but the ideal laboratory where all the tests of the modelling of the pulsation
on horizontal branch stars can be performed. 
The persistence of the current pulsation period in the low status  was
established by the new $T_{\rm max}$ values obtained in 2015 \citep{konkoly}. 
The survey with the  
 ``Very Tiny Telescopes" \citep[VTTs; ][]{itself}  has continued in 2016 and 2017. Figure~\ref{storia} shows
the development of the pulsation period in the last three years: the new $T_{\rm max}$ values 
indicate that the pulsation is still in the short-period status. The pulsation period
is  actually the shortest never experienced by the star.

\section{The Blazhko period of RR Lyr}
We also determined the Blazhko period since 1910 to verify how it follows
the changes in the pulsation period \citep{itself}.
We could infer that  the variations of the pulsation
and  Blazhko periods are  completely decoupled, since the Blazhko period
had just one sudden decrease from 40.8~d to 39.0~d in 1975.

\begin{figure}
\begin{minipage}{1.0\textwidth}
\centering
\includegraphics[width=\textwidth]{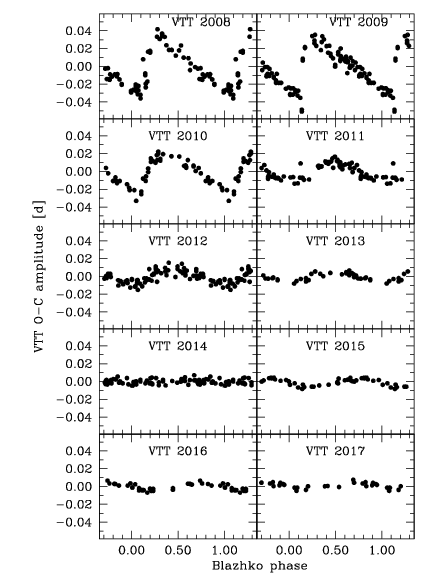}
\caption{The behaviour of the amplitude  of the Blazhko effect of RR Lyr. The  new determinations 
in 2016 and 2017 are shown in the bottom panels. 
}
\label{vtt}
\end{minipage}
\end{figure}

\begin{figure}
\begin{minipage}{1.0\textwidth}
\centering
\includegraphics[width=\textwidth]{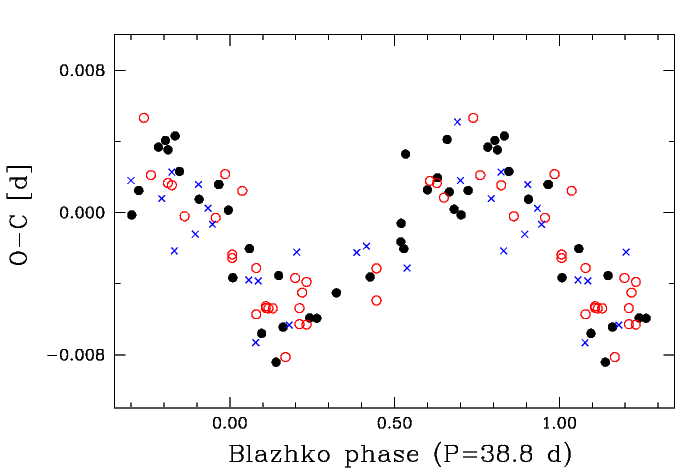}
\caption{The O-C values obtained in 2015 (filled black circles), 2016 (empty red circles) and
2017 (blue squares)  folded with the Blazhko period P$_B$=38.8~d. 
}
\label{vtt17}
\end{minipage}
\end{figure}

We remind the reader that the measurements  obtained by the VTTs and by {\it Kepler} have also been
extensively used to record the monotonic long-term decrease in the amplitude of the Blazhko effect
\citep{itself}.
Such a decrease resulted in an almost final damping  in 2014. The subsequent $T_{\rm max}$ values  
  collected with the VTTs in 2015 showed
a slight increase in the amplitude of the O-C values \citep{konkoly}.  The new values collected in
2016-17 confirm that the Blazhko effect has resumed at a measurable amplitude (Fig.~\ref{vtt}). 
However, the amplitudes in 2015 and 2016-17 look very similar (Fig.~\ref{vtt17}) and hence the
effect is still weak.

\section{Hunting for the first RR Lyr star in a binary system}
We do not know any binary system in which one of the component
is an RR Lyr variable. Searching for such a system
%The senarch for an RR Lyr star belonging to  a binary system 
is not only an observational exercise. Our knowledge on the masses of RR Lyr stars is
based on evolutionary and pulsation models. A direct measure is still lacking.
As in the case of many fields of stellar astronomy, binary stars can provide the laboratory where
this measure can be extracted. 
%The direct comparison among dynamical, pulsational,
%and evolutionary masses will allow us to verify the hypotheses and assumptions
%of the theoretical models.

The {\it Kepler} field includes many RR Lyr variables, new and old
%both already known and new discoveries
\citep{benko}. The almost continuous light curves supply a testbed for the careful
analysis of the O-C values in search for cyclic patterns due to the light-time effect.
Such an analysis is particularly powerful in the case of {\it Kepler} stars since
they have been observed for 4-years, investigating a wide range of orbital periods.
A very promising candidate has been recently proposed: KIC~2831097 \citep{sodor}.
It is a first-overtone RR  Lyr variable 
with a period $P$=0.337~d and a light amplitude $\Delta K_p$=0.40~mag. 
%Like in other RR Lyr variables, a few additional modes have also been detected. 
%KIC~2831097 also shows a long-term
%irregular modulation of about 47~d resembling the Blazhko effect.
\begin{figure}
\begin{minipage}{1.0\textwidth}
\centering
\includegraphics[width=\textwidth]{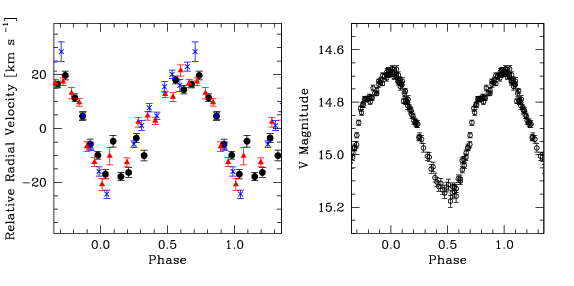}
\caption{
{\it Left panel:} the radial velocity curve over the pulsation
period obtained with FIES in August 2017. 
%Mean value subtracted from the data.
%Error bars ($\pm1\sigma$) are in the range 1.5-3.5~km\,s$^{-1}$.
Different symbols for different nights.
{\it Right panel:} the light curve from simultaneous
 CCD photometry.
%obtained by M.~Correa (Agrupaci\'o Astron\`omica de Sabadell, Spain) with
% a 30-cm telescope in $V$-light.}
}
\label{fies}
\end{minipage}
\end{figure}

The O-C curve shows two overlapping  effects: a linear decrease of the pulsation
period, as observed in many RR Lyr variables \citep{geos} and a
well-defined periodicity of 753~d with  a full-amplitude of 0.04~d.
This periodicity %is unique in the observational scenario of RR Lyr stars and
seems a well-established fact since it has been followed for almost two full cycles
and repeats itself in a very regular way \citep[Fig.~2, top panel, in ][]{sodor}. 
We can look
at KIC~2831097 as the currently best target for hunting  the
first RR Lyr star in a binary system. 

In our collaboration we obtained a clear radial-velocity curve
from the H$\alpha$ line 
%(Fig.~1, left panel; the mean systemic
%velocity has been subtracted to protect our result) 
by using the FIES instrument
mounted at the Nordic Optical Telescope (Roque de Los Muchachos, La Palma, Canary Islands,
Spain).
Figure~\ref{fies} (left panel) shows the radial velocity obtained from the H$\alpha$ line,
folded over the pulsation period;
the mean systemic velocity has been subtracted. 
The spectra were taken in August 2017, near one of the quadratures. 
A similar campaign is planned in May 2018 to obtain
the pulsation radial velocity at the other quadrature.

The GEOS collaboration also ensured the simultaneous photometric coverage:
Fig.~\ref{fies} (right panel) shows the $V$ light-curve obtained 
%Error bars ($\pm1\sigma$) are in the range 1.5-3.5~km\,s$^{-1}$.
%Different symbols for different nights.
 by the Agrupaci\'o Astron\`omica de Sabadell (Spain) with
 a Meade 12\verb+"+ telescope equipped with a CCD Moravian G2 1600. 
%in $V$-light.  Meade 12" f/ 5,8 CCD Moravian G2 1600

%We also measured a full-amplitude
%of 50~km~s$^{-1}$ of the pulsational radial-velocity curve. Therefore, we can confirm that the
%expected
%difference between the systemic velocities is much larger than the pulsational amplitude, making
%the keplerian motion easily to be detected after the second run requested here.

\section{Searching for binary RR Lyr stars in the GEOS survey}

Since 2004, the GEOS survey uses TAROT telescopes to measure the O-Cs of $T_{\rm max}$ values  of hundreds 
of galactic RRLyr stars \citep{geos}. The O-C accuracy is about 3 minutes and some stars 
are observed many times per year. As a consequence, such a density of O-C values leads to identify 
new variations with periods longer than ten years. They could be due to a companion of 
the RR Lyr star, thus provoking   a light-time effect. The baseline of only 13 years of the 
GEOS survey introduces a bias leading to find suspect companions nearest than 20 AU from their RR~Lyr star. 
It remains difficult to discriminate if variations are due to the light-time effect or to a very long
Blazhko period from photometric data only. Taking into account the large distance of the stars from the Earth, 
the apparent angular separation is typically lower than 10~mas, making it very challenging for direct 
confirmation by using adaptive optics. We plan to continue the TAROT observations to confirm the 
suspected  periodicities  and to find other stars with longer periods.

%\begin{figure}
%\begin{minipage}{0.35\textwidth}
%\centering
%\includegraphics[width=\textwidth]{PorettiFig1.PNG}
%\caption{Graphical representation of the projection factor $p$ between the
%true pulsational velocity $V_{\rm puls}$ of the star and the radial velocity $V_{\rm rad}$
%measured by the observer.}
%\label{fig:p0}
%\end{minipage}
%\quad

%\begin{figure}
%\centering
%\includegraphics[width=0.75\textwidth]{PorettiFig3.pdf}
%\caption{Graphical representation of the decomposition of the projection
%factor $p$: atmospheric gradient (line ``1"), geometric (shift ``2"), extrapolation
%to the photosphere (``3"), and correction for the radial velocity of the gas
%(shift ``4").}
%\label{proj}
%\end{figure}

%\begin{figure}
%\includegraphics[width=\textwidth]{PorettiFig4.pdf}
%\caption{HARPS-N variation of the amplitude of the radial velocity curve $\Delta RV_{\mathrm{c}}$ 
%in function of the line depth $D$.}
%\label{depth}
%\end{figure}

\section{Conclusions}
The VTT monitoring showed that the pulsation of RR~Lyr has continued in its short-period status
in 2016 and 2017.
Moreover, the VTT still measured a weak amplitude of the Blazhko effect, but larger than in 2014.
Both facts confirm the importance of continuing the observation of this star, since
its behaviour through the ages can bring important information on the evolution
of horizontal-branch stars.  

As a new topic in the GEOS research, we investigated the TAROT database to identify RR~Lyr stars in
binary systems. At the moment,  KIC~2831097 remains the best candidate for such a system. We
started an observational project to obtain the radial velocity curves of this star at the predicted
quadratures. The first campaign was successful, while the second is programmed 
in 2018.

\acknowledgements{
EP and MR acknowledge financial support from the PRIN-INAF 2014.
}

\bibliographystyle{ptapap}
\bibliography{EPoretti}

\end{document}